\documentstyle[prl,aps,multicol,epsfig]{revtex}
\begin{document} 
\draft 
\title{Excitation spectrum of the homogeneous spin liquid}

\author{R. Eder}
\address{Institut f\"ur Theoretische Physik, Universit\"at W\"urzburg,
Am Hubland,  97074 W\"urzburg, Germany}
\date{\today}
\maketitle

\begin{abstract}
We discuss the excitation spectrum of a disordered, isotropic and 
translationally invariant spin state in the 2D Heisenberg antiferromagnet. 
The starting point is the nearest-neighbor RVB state which plays the role of 
the vacuum of the theory, in a similar sense as the N\'eel state is the vacuum
for antiferromagnetic spin wave theory. We discuss the elementary excitations 
of this state and show that these are not Fermionic spin-1/2 `spinons' but 
spin-1 excited dimers which must be modeled by bond Bosons. We derive an 
effective Hamiltonian describing the excited dimers which is formally 
analogous to spin wave theory. Condensation of the bond-Bosons at zero 
temperature into the state with momentum $(\pi,\pi)$ is shown to be equivalent 
to antiferromagnetic ordering. The latter is a key ingredient for a 
microscopic interpretation of Zhang's SO(5) theory of cuprate 
superconductivity.
\end{abstract} 
\begin{multicols}{2}
\section{Introduction}
High-temperature superconductivity occurs in a state
which is is frequently referred to as an `RVB spin liquid'. This state has no
magnetic order, but strong short range antiferromagnetic correlations.
Undoubtedly the strong repulsion between electrons, which turns
the system into a charge-transfer insulator at half-filling
persists in the doped case, so that a description in terms of free-electron 
like Slater-determinants is not really adequate. The problem then
is, how to describe such a state theoretically.
Despite its frequently being referred to in the literature,
the RVB spin liquid is a rather elusive concept.
For example the precise nature of its ground state and low lying elementary 
excitations is not known to any degree of certainty.
In the following we want to address this problem by studying
a disordered state for the $2D$ Heisenberg antiferromagnet
\[
H = J \sum_{\langle i,j\rangle} \bbox{S}_i \cdot \bbox{S}_j
\]
on an 2D square lattice. Here $\bbox{S}_i$ denotes a spin 1/2
operator on site $i$. One might expect that this is a kind of stepping
stone also for the doped case, in that the elementary excitations
of the undoped spin liquid persist to some degree also for finite
doping.\\
Perhaps the best-defined RVB spin liquid is the
nearest neighbor RVB 
state\cite{Kivelson,Sutherland,Kohmoto,Fradkin} 
- at least this wave function can be written down in compact form.
We define the singlet generation operator on the bond  $i,j$ 
\begin{equation}
s_{i,j}^\dagger = \frac{1}{\sqrt{2}}
(\;\hat{c}_{i,\uparrow}^\dagger
 \hat{c}_{j,\downarrow}^\dagger -
\hat{c}_{i,\downarrow}^\dagger
\hat{c}_{j,\uparrow}^\dagger\;),
\label{singlet}
\end{equation}
where $\hat{c}_{i,\sigma}^\dagger$$=$$c_{i,\sigma}^\dagger
c_{i,\bar{\sigma}}^{} c_{i,\bar{\sigma}}^\dagger$, are the
constrained Fermion operators, which do not allow
the creation of a second electron on an already singly occupied site.
Introducing the operator
\begin{equation}
S = \sum_i ( s_{i,i+\hat{x}}^\dagger +  s_{i,i+\hat{y}}^\dagger),
\label{sdef}
\end{equation}
where e.g. $i+\hat{x}$ denotes the nearest neighbor of site $i$
in $x$-direction,
the nearest neighbor RVB state on a 2D square lattice with $2N$ sites
can be written as
\begin{equation}
|RVB\rangle = \frac{1}{\sqrt{n}} 
\frac{S^{N}}{N!} |0\rangle,
\label{RVB}
\end{equation}
where $n$ is a normalization factor.
It corresponds to a superposition of all states which can be obtained by
covering the plane compactly with nearest neighbor
singlets, all with equal phase.
Covering the plane with singlets is equivalent to covering it with
dimers, a well-known problem from statistical mechanics\cite{Kasteleyn}.
We can therefore rewrite the state as
\begin{equation}
|RVB\rangle 
=\frac{1}{\sqrt{n}}  \sum_{a} |\psi_a\rangle,
\label{RVB1}
\end{equation}
where $a$ denotes a dimer covering of the lattice and
$|\psi_a\rangle$ the state obtained by putting
singlets onto the dimers of $a$.\\
In the following, we want to examine the problem
of the possible elementary excitations of such a singlet background,
and set up an effective Hamiltonian governing their dynamics.
The idea of `expanding' around a suitably chosen vacuum state
is realized in simplest form in linear spin wave theory.
The general line of thought here is quite analogous to
linear spin wave theory, with the sole exception that
the role of the vacuum (which determines the symmetries
of the ground state) is played by the `singlet soup' (\ref{RVB})
instead of the N\'eel state. 
A similar approach has 
previously been applied to dimerized planar Heisenberg-type
models\cite{Sachdev,singh}
to spin ladders\cite{Gopalan,lad}, to strongly coupled 
Heisenberg planes\cite{chubukov,Kotov}, and to Spin-Peierls-like
spin chains\cite{wang,brenig}. An example where
the fluctuations are Fermionic rather than Bosonic in nature
is provided by the Kondo lattice\cite{Oana}.
The main difference as compared to the present work is that in all of 
these works a rather unique and simple dimer covering of the system is
given by the topology or the form of the Hamiltonian - the 
complications that arise from the use 
of a disordered `singlet soup' such as (\ref{RVB}) then can be avoided.\\
While the technical complications arising from the use of a dimer basis are
considerable, this approach also has some major advantages:
in a site basis it is virtually impossible to even write down
a disordered spin state, because one has to deal with the spin degeneracy
on each single site. 
The calculation only becomes feasible if this site-degeneracy is lifted,
for example by assuming strict N\'eel order as in spin wave theory.
On the other hand, the degeneracy is automatically taken care of
in the dimer basis, because two interacting sites do have a unique ground
state. A further considerable advantage of the dimer basis is,
that it is easily enlarged by hole pairs on nearest neighbors, so
as to describe a superconducting state. Indeed, as will be seen
below, the present description of the antiferromagnetic phase
most naturally can be generalized to comprise also a superconducting
phase, thereby providing a very simple microscopical picture for 
the SO(5) rotations which smoothly connect antiferromagnetic
and superconducting phase in Zhang's
theory\cite{zhang} of cuprate superconductors.
\section{Elementary Excitations of a `Singlet Soup'}
\begin{figure}
\epsfxsize=7.0cm
\vspace{-0.0cm}
\hspace{0.5cm}\epsfig{file=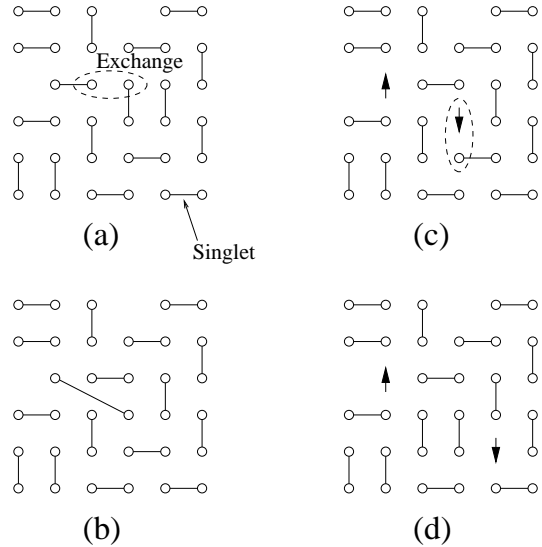,width=7.0cm,angle=0.0}
\vspace{0.5cm}
\narrowtext
\caption[]{An incorrect picture for fluctuations in a
singlet background.}
\label{fig1} 
\end{figure}

The nearest neighbor-RVB state (\ref{RVB1}) has the
symmetry properties expected for a homogeneous spin liquid:
it is isotropic, translationally invariant, is an exact spin singlet
and has no magnetic order.
On the other hand, just as the N\'eel state, it is not an eigenstate of $H$.
If we take one singlet configuration
$|\psi_a\rangle$ and act with the exchange term
on a bond connecting two different singlets 
(see Figure \ref{fig1}a) we can create a state which
no longer can 
be represented
as a superposition of only nearest neighbor singlets. Such
a state therefore represents a kind of fluctuation and
as a first step we need to understand the character of these fluctuations.
It might appear\cite{LiangDoucotAnderson} 
that the energetically most favorable 
fluctuation is the state shown in Figure \ref{fig1}b:
two nearest neighbor singlets are transformed into
a configuration with only one nearest neighbor singlet
and a second singlet connecting more distant sites.
Nominally the energy increases by only $3J/4$ in this transition, 
because the only change is the loss of
one nearest neighbor singlet. Because singlet and triplet are
degenerate for sites which are not connected by an exchange bond,
we might as well consider the two spins connected by the `long singlet' 
as being unpaired (see Figure \ref{fig1}c).
The transition from Figure \ref{fig1} a$\rightarrow$b could 
thus be viewed as pair creation of two unpaired spins.
Next, by acting with the exchange term on a bond which connects a 
dangling spin to another singlet (see Figure \ref{fig1}c), 
we can recouple the
spins so as to form a new singlet and leave one of the formerly
paired spins unpaired
(see Figure \ref{fig1}d). This process corresponds to a
propagation 
\noindent 
of the dangling spin. We would thus arrive at the
conclusion that the fluctuations out of the 
nearest neighbor RVB state are unpaired spins,
which carry a spin of $1/2$ and consequently must obey Fermi statistics.
Clearly, these excitations should be identified with the ominous `spinons'.\\
Further reasoning shows, however, that the line of thought leading to
the introduction of the `spinons' misses a small but crucial detail.
The first reason is that the
state in Figure \ref{fig1}b is not orthogonal to the
vacuum, 
\begin{figure}
\epsfxsize=4.0cm
\vspace{-0.0cm}
\hspace{1.5cm}\epsfig{file=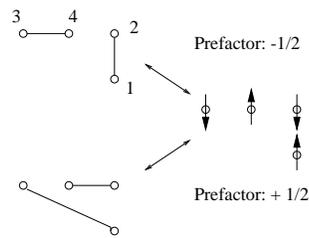,width=4cm,angle=0.0}
\vspace{0.5cm}
\narrowtext
\caption[]{Placing singlets in either of the two ways
shown on the left produces the
spin configuration on the right with the indicated
prefactors. This gives a contribution of $-1/4$ to the
overlap, an equal contribution comes from the
spin reversed configuration.}
\label{fig2} 
\end{figure}
\noindent 
and thus cannot represent a true fluctuation.
More precisely, if we introduce (see Figures \ref{fig1} and \ref{fig2}))
\begin{eqnarray*}
|a\rangle &=&  s_{1,2}^\dagger s_{3,4}^\dagger |0\rangle,
\nonumber \\
|b\rangle &=&  s_{1,3}^\dagger s_{4,2}^\dagger |0\rangle,
\end{eqnarray*}
it is straightforward to see (Figure \ref{fig2}) that
$\langle a |b \rangle$$=$$-\frac{1}{2}$,
in other words: after the `transition'
Figure \ref{fig1}a$\rightarrow$b we remain in the original
state, Figure \ref{fig1}a, with a probability of $25$\%.
The problem of non-orthogonality
is not restricted to the first step
in Figure \ref{fig1}: the states
Fig. \ref{fig1}c and \ref{fig1}d have an overlap of $1/2$,
and this generalizes to any two states which differ by one
hopping process of a `spinon'. The non-orthogonality problem
thus is omnipresent and severe.\\
Let us therefore return to the first step,
Figure \ref{fig1} a$\rightarrow$b, and consider how we can
remedy the problem. The most natural way
to proceed is to form the orthogonal complement
\[
|b'\rangle = |b\rangle - \langle a|b\rangle\; |a\rangle,
\]
so as to see `what is really new' in the state $|b\rangle$.
A straightforward computation shows that after
normalization to unity the orthogonal complement is
\begin{equation}
|b'\rangle = \frac{1}{\sqrt{3}}
\sum_{\alpha=x,y,z}  t_{12,\alpha}^\dagger t_{34,\alpha}^\dagger |0\rangle.
\label{compl}
\end{equation}
Here we have introduced the operators\cite{chubu,Sachdev}
\begin{eqnarray}
t_{ij,x}^\dagger &=& 
\frac{-1}{\sqrt{2}}(
\hat{c}_{i,\uparrow}^\dagger \hat{c}_{j,\uparrow}^\dagger
-\hat{c}_{i,\downarrow}^\dagger \hat{c}_{j,\downarrow}^\dagger),
\nonumber \\
t_{ij,y}^\dagger &=& 
\frac{i}{\sqrt{2}}(
\hat{c}_{i,\uparrow}^\dagger \hat{c}_{j,\uparrow}^\dagger +
\hat{c}_{i,\downarrow}^\dagger \hat{c}_{j,\downarrow}^\dagger),
\nonumber \\
t_{ij,z}^\dagger &=& \frac{1}{\sqrt{2}}
(\hat{c}_{i,\uparrow}^\dagger \hat{c}_{j,\downarrow}^\dagger
+ \hat{c}_{i,\downarrow}^\dagger \hat{c}_{j,\uparrow}^\dagger),
\label{triplets}
\end{eqnarray}
which create the three components of the {\em triplet}
on the bond $(i,j)$. 
We arrive at the conclusion that the true fluctuation
out of the nearest neighbor singlet background is
not the creation of two Fermionic `spinons', but rather the 
creation of two excited dimers, which carry
a spin of $1$ and consequently should be modeled by 
bond-Bosons\cite{chubu,Sachdev}. 
The further evolution of the
created triplets then is quite obvious (see Figure \ref{fig3})
(but completely different from that of the `spinons'): by
exchange along 
bonds connecting the triplets with neighboring
singlets the triplets can de-excite
while simultaneously the singlet turns into a triplet - this process, which
is very much reminiscent of the propagation of a Frenkel-type
exciton corresponds to the propagation of the
excited dimer. Note that unlike the `spinon'
states in Figure \ref{fig1}, all different
states in Figure \ref{fig3}
are mutually rigorously orthogonal. As a matter
of fact there are problems with non-orthogonalities also for the
triplet states - these are `inherited' from the original
nearest neighbor RVB state.
They will be discussed in detail below and be shown to
be much less severe than those for the `spinon' states.
Their main effect is to replace the simple excited dimer by a
more delocalized object, which resonates between different
orientations within a limited spatial region.
\begin{figure}
\epsfxsize=7.0cm
\vspace{-0.0cm}
\hspace{0.5cm}\epsfig{file=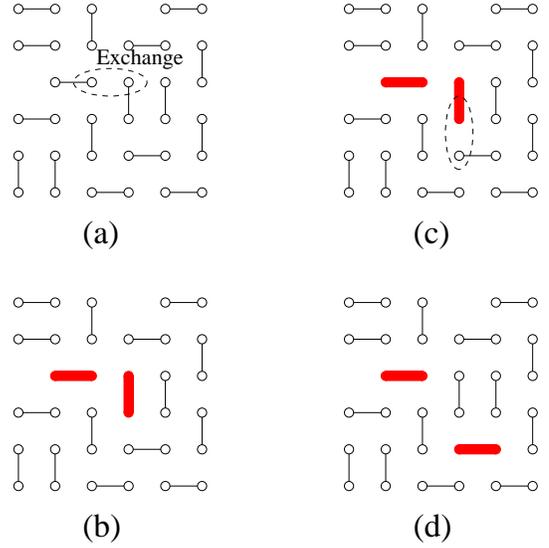,width=7.0cm,angle=0.0}
\vspace{0.5cm}
\narrowtext
\caption[]{A more correct picture for fluctuations
in a singlet background.}
\label{fig3} 
\end{figure}
\noindent 
To be more precise, we now discuss the action of the
Heisenberg exchange on all possible configurations
of nearest neighbor singlets and triplets.
Consider two {\em nearest neighbor bonds}
$(i,j)$ and $(i',j')$, and assume that they are connected by
a single bond $(i,i')$. Denoting the
Heisenberg exchange along the latter bond by
$h_{i,i'}$ we have:
\begin{eqnarray}
h_{i,i'} s_{i,j}^\dagger s_{i',j'}^\dagger
&=& \frac{J}{4} \bbox{t}_{i,j}^\dagger \cdot \bbox{t}_{i',j'}^\dagger,
\nonumber \\
h_{i,i'} s_{i,j}^\dagger \bbox{t}_{i',j'}^\dagger
&=& \frac{J}{4} \bbox{t}_{i,j}^\dagger s_{i',j'}^\dagger
-\frac{iJ}{4} \bbox{t}_{i,j}^\dagger \times \bbox{t}_{i',j'}^\dagger,
\nonumber \\
h_{i,i'} t_{i,j,\alpha}^\dagger t_{i',j',\alpha}^\dagger
&=& \frac{J}{4}  s_{i,j}^\dagger s_{i',j'}^\dagger
\nonumber \\
&\;&\;\;\;\;\;
- \frac{J}{4} (\bbox{t}_{i,j}^\dagger \cdot \bbox{t}_{i',j'}^\dagger
-  t_{i,j,\alpha}^\dagger t_{i',j',\alpha}^\dagger),
\nonumber \\
h_{i,i'} t_{i,j,\alpha}^\dagger t_{i',j',\beta}^\dagger
&=& \frac{J}{4} t_{i,j,\beta}^\dagger t_{i',j',\alpha}^\dagger
\nonumber \\
&\;&\;\;\;\;\;- \frac{iJ\epsilon_{\alpha\beta\gamma}}{4}
( t_{i,j,\gamma}^\dagger s_{i',j'}^\dagger -
s_{i,j}^\dagger t_{i',j',\gamma}^\dagger ).
\nonumber \\
\label{st}
\end{eqnarray}
These equations show that
if we start out from states containing nearest neighbor singlets
or triplets on the l.h.s., the exchange term only
produces states which again consist of nearest neighbor
singlets or triplets on the
r.h.s (had the two bonds been connected by exchange
along another bond, $(j,i')$, $(i,j')$ or $(j,j')$, we would have
obtained the same equations with the sole difference that
in some cases the prefactors change their sign).
This proves rigorously that acting with an arbitrarily high power of the
Hamiltonian onto the nearest neighbor RVB state produces only states
which can be built up from {\em nearest neighbor} singlets or triplets.\\
This `theorem' in fact holds true in a more general sense:
on a single dimer $m$, the $4$ states $s_m^\dagger$ and $\bbox{t}_m^\dagger$
do form a complete basis\cite{Sachdev}. Thus, if we use a fixed
dimer covering $a$, the set of states obtained by placing singlets and triplets
on the dimers of $a$ form a complete basis of the Hilbert space.
Adding up such states obtained from {\em all possible} dimer coverings
then clearly produces a highly overcomplete basis of the Hilbert
space, which therefore must automatically include
states with singlets of arbitrary length.
It follows that all states with
longer-range singlets also can be represented as superpositions
of states which are composed exclusively from nearest-neighbor singlets and
triplets. These states are therefore redundant, and
if we formulate a self-consistent
theory in terms of nearest-neighbor singlets and
triplets, we have automatically included these longer-ranged
singlets. The fact that we are using a nearest-neighbor
RVB state as the starting point for constructing
our theory therefore means no loss of generality and in particular
does by no means imply that we are considering only states
with only very short-ranged antiferromagnetic correlations. In fact,
it will be shown below that one can construct even states with
infinite-range antiferromagnetic
order by using exclusively nearest neighbor singlets and triplets.\\
The preceding considerations suggest that
we should model the excitation spectrum of the
nearest neighbor singlet vacuum by Bosonic
excitations, which approximately correspond to excited dimers.
Assuming that the bonds in the system
have been labeled in some way, we denote
the triplet operator on bond $i$ by
$\bbox{t}_i^\dagger$. Then, we introduce the following basis states
\begin{eqnarray}
|\Psi_{i_1\alpha_1, i_2\alpha_2,\dots i_m\alpha_m}\rangle
&=&
\frac{1}{\sqrt{ n(i_1\alpha_1, i_2\alpha_2,\dots i_m\alpha_m)}}
\nonumber \\
&\;&\;\;\;
\frac{S^{N-m}}
{(N-m)!}
\prod_{\nu=1}^{m}
t_{i_\nu,\alpha_\nu}^\dagger |0\rangle,
\label{basis}
\end{eqnarray}
where $ n(i_1\alpha_1, i_2\alpha_2,\dots i_m\alpha_m)$ is a
normalization factor.
They describe a certain number ($m$) of triplets
which are `immersed into the singlet soup'. Thereby the singlets
fill the space in between the triplets compactly in all possible ways.
All states which can be generated by pair creation
and propagation of triplet bonds (such as the ones shown in Figure
\ref{fig3}) can be represented in this way.
We next consider the triplets as Boson-like elementary
excitations of the singlet-background, in precisely
the same way as misaligned
spins are considered as Bosonic excitations in a `N\'eel background'
in antiferromagnetic spin wave theory.
Re-interpreting the state
\[
|\psi_{i_1\alpha_1, i_2\alpha_2,\dots i_n\alpha_m}\rangle
\rightarrow
\prod_{\nu=1}^{m}
\tau_{i_\nu,\alpha_\nu}^\dagger |0\rangle,
\]
where the $\tau_{i_\nu,\alpha_\nu}^\dagger$ represent
Boson operators, we may expect to describe the
dynamics of these Bosons by a Hamiltonian of the form\cite{remark}
\begin{eqnarray}
H &=& J \sum_i \bbox{\tau}_i^\dagger \cdot  \bbox{\tau}_i^{}
+ \sum_{i,j} (\Delta_{ij} \bbox{\tau}_i^\dagger \cdot 
\bbox{\tau}_j^{\dagger} + H.c.)
\nonumber \\
&\;&\;\;\;\;\;\;\;\;\;\;\;\;\;\;\;\;\;\;\;\;\;\;\;\;\;\;\;\;\;
+ \sum_{i,j} \epsilon_{ij}\; \bbox{\tau}_i^\dagger \cdot \bbox{\tau}_j^{},
\label{heff}
\end{eqnarray}
where we have grouped the three triplet components into
a 3-vector $\bbox{\tau}$ so as to stress manifest rotation invariance.
The first term in (\ref{heff}) corresponds to the
energy of formation of the triplets,
the second term describes pair creation processes as 
in Figure \ref{fig3}a $\rightarrow$ \ref{fig3}b, and the third term
accounts for the propagation of the triplets, see
Figure \ref{fig3}c $\rightarrow$ \ref{fig3}d.
The matrix elements $\epsilon_{ij}$
and $\Delta_{ij}$ should be obtained by
computing matrix elements of the Heisenberg
Hamiltonian $H$ between the corresponding
states (\ref{basis}). Of course one thereby has to assume that
for example the matrix element for a triplet jumping from
bond $m$ to bond $n$ does not depend significantly
on the positions of the other triplets - otherwise
a description in terms of a single-particle like Hamiltonian
would not be feasible.
As is the case in spin wave theory, the $\tau$-Bosons have to obey 
a hard-core constraint, and in fact presence of one Boson prohibits
the presence of another Boson not only on the same bond,
but also on all bonds which share a site with the original one.\\
In the following, we will first
study the problem of a single
excited dimer in the singlet background, in other words we want to
compute the `bare' Boson dispersion $\epsilon(\bbox{k})$ in
(\ref{heff}). As our basis states we consequently choose (suppressing the
$x$ $y$ or $z$ spin-index of the triplet):
\begin{equation}
|i,\alpha\rangle
=  \frac{1}{\sqrt{n_1}}
\frac{S^{N-1}}{(N-1)!} \; t_{i, i+\hat{\alpha}}^\dagger 
\;|0\rangle,
\label{1bond}
\end{equation}
where $\alpha$$=$$x,y$ denotes the
direction of the bond in real space, and $n_1$ a normalization factor.
In this state one triplet is put onto the bond $(i,i+\hat{\alpha})$
and the remainder of the lattice is covered compactly
by singlets in all possible ways. 
Next, we introduce the Fourier transforms
\begin{equation}
|\bbox{k},\alpha\rangle
= \frac{e^{i k_\alpha/2}}{\sqrt{2N}}
\sum_j  
|j, \alpha\rangle  e^{i \bbox{k} \cdot \bbox{R}_j}.
\label{fou1}
\end{equation}
In the Hilbert space of bond-Bosons, this state
would be denoted by $\tilde{\tau}_{\bbox{k},\alpha}^\dagger |0\rangle$.
The procedure to be followed then is like this: in a first step we 
compute the $2\times 2$ overlap matrix 
$N(\bbox{k})$$=$$\langle \tilde{\tau}_{\bbox{k},\alpha}^{}
\tilde{\tau}_{\bbox{k},\alpha'}^\dagger\rangle$
and diagonalize it. Denoting the resulting eigenvectors and 
eigenvalues by $\bbox{e}_\nu$ and $\lambda_\nu$, the states
\begin{equation}
\tau_{\bbox{k},\nu}^\dagger |0\rangle =
\frac{1}{\sqrt{\lambda_\nu}} \sum_{\alpha =x,y}
\bbox{e}_{\nu,\alpha}
\tilde{\tau}_{\bbox{k},\alpha}^\dagger|0\rangle
\label{ortho}
\end{equation}
form an orthonormal basis set and hence can serve as effective
single particle orbitals with momentum $\bbox{k}$.
Since the Boson operators which correspond to the original
triplets obey\cite{stollie} $[\tilde{\tau}_{\bbox{k},\alpha}^{},
\tilde{\tau}_{\bbox{k},\beta}^\dagger] =
N_{\alpha,\beta}$, the operators $\tau_{\bbox{k},\nu}$
obey the canonical commutation relations for Boson operators:
$[\tau_{\bbox{k},\nu^{}},\tau_{\bbox{k},\nu'}^\dagger]=\delta_{\nu,\nu'}$.
They describe a triplet-like excitation which oscillates between
$x$ and $y$-directed bonds within a certain spatial region
whose extent is determined by the range of
the real space overlap integrals $\langle i, \alpha| j, \alpha'\rangle$.\\
Next, we set up the $2\times 2$ Hamilton matrix
$H(\bbox{k})$$=$$\langle \tau_{\bbox{k},\nu}^{}|H
|\tau_{ \bbox{k},\nu'}^\dagger\rangle$,
which in turn requires knowledge of the real space matrix elements
$\langle i, \alpha|H| j, \alpha'\rangle$.
Diagonalizing $H(\bbox{k})-E_0N(\bbox{k})$, 
where $E_0$ denotes the expectation value of $H$ in the
`background' state (\ref{RVB1}), we obtain the desired dispersion relation 
of a single triplet Boson. The pair creation matrix element is obtained
in an analogous way.\\
This procedure in fact is neither new nor 
unconventional: a completely analogous construction
is performed e.g. in the construction of the $t-J$ model\cite{ZhangRice}, 
which
describes the dynamics of the (non-orthogonal) Zhang-Rice singlets on the
different plaquettes of the CuO$_2$ plane. The only difference is that
here we have two different 
nonorthogonal objects (Bosons on bonds in $x$ and $y$-direction)
per unit cell, whereas it was only a single Zhang-Rice singlet/unit cell
in the case of the CuO$_2$ plane. Apart from that the construction is
precisely the same.\\
In the next three sections we will calculate the
dispersion relation, the pair creation matrix element, and 
discuss how these matrix elements depend on the density of
triplets. Readers which are not interested in these more technical
parts are advised to proceed to section VI.
\section{Propagation of a single triplet}
\begin{figure}
\epsfxsize=8.0cm
\vspace{-0.5cm}
\hspace{-0.5cm}\epsfig{file=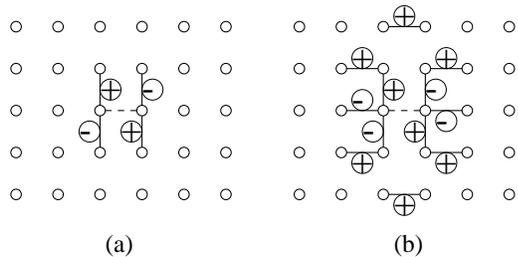,width=8.0cm,angle=0.0}
\vspace{0.5cm}
\narrowtext
\caption[]{Pairs of bonds which are connected
by $L$$=$$2$ loops (a) and
$$L$$=$$3$$ loops (b). Bond $i$ is kept fixed (dashed line),
bond $j$ (full line) is labeled by $(-1)^{\sigma(u)}$.}
\label{fig4} 
\end{figure}

\begin{figure}
\epsfxsize=5.0cm
\vspace{-0.5cm}
\hspace{1.5cm}\epsfig{file=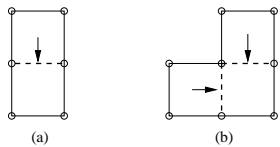,width=5.0cm,angle=0.0}
\vspace{0.5cm}
\narrowtext
\caption[]{`Bridging bonds' (dashed lines) in the $L=3$ loop (a) and in an
$L=4$ loop (b).}
\label{fig5} 
\end{figure}

\begin{figure}
\epsfxsize=3.0cm
\vspace{-0.5cm}
\hspace{-0.5cm}\epsfig{file=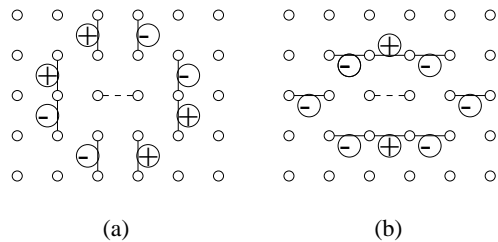,width=8.0cm,angle=0.0}
\vspace{0.0cm}
\narrowtext
\caption[]{The sign of the hopping integral
from the dashed bond $m$ (dashed) to the
indicated bond $n$.}
\label{fig6} 
\end{figure}

\begin{figure}
\epsfxsize=4.0cm
\vspace{-0.0cm}
\hspace{0cm}\epsfig{file=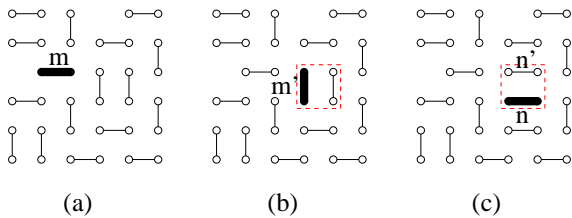,width=8cm,angle=0.0}
\vspace{0.5cm}
\narrowtext
\caption[]{By application of $H$ a triplet can hop from bond
$m\rightarrow m'$ (see (a)$\rightarrow$(b)) and then
overlap with the triplet in the final state on bond $n$
along the indicated $L=2$-loop (see (b)$\rightarrow$(c)).
This process gives a nonvanishing hopping matrix element
from bond $m\rightarrow n$.}
\label{fig7} 
\end{figure}

To carry out our program we need to
compute the real-space matrix elements
$\langle i, \alpha| j, \beta\rangle$ and
$\langle i, \alpha|H| j, \beta\rangle$. In doing so 
the concept of a loop covering of the plane\cite{Sutherland}
is of crucial importance.
For two dimer coverings $a$ and $b$ the loop covering
$c=a+b$ is obtained by drawing $a$ and $b$ `on top of each other'
(see for example Figure 1 in Ref. \cite{Sutherland}).
This produces a covering of the plane by closed loops $u$,
each of even length $2L(u)$ (note that in the following
we always measure the length of a loop $L(u)$ `in units of dimers').
Let us now consider each loop $u$ as an isolated 1D ring with
$2L(u)$ sites. We assume that the sites along the ring are labeled such
that the dimer covering $a$ corresponds to the state
$|a\rangle=\prod_i s_{i,i+1}^\dagger |0\rangle$ - the dimer covering $b$ then
must correspond to $|b\rangle=\prod_i s_{i+1,i+2}^\dagger |0\rangle$.
Expanding the products we get $2^{L(u)}$ different spin states
from each covering, and there are precisely $2$ spin states
which show up in both $|a\rangle$ and $|b\rangle$
namely the two possible N\'eel states. We thus have
$\langle a|b\rangle=2/2^{L(u)}$. 
The same holds true for any other loop, whence,
using $\sum_{u\in a+b}L(u) = N$, we find
the scalar product of the two singlet distributions\cite{Sutherland}
\[
\langle \psi_a | \psi_{b} \rangle = 2^{P(a+b)-N},
\]
where $P(c)$ is the total number of loops
in the loop covering $c$.\\
Let us now assume that the singlets on the bond $m$ in
$|\psi_a\rangle$ and on the bond $n$ in $|\psi_{b}\rangle$
have been replaced
by a $z$-like triplet (due to the explicit rotational invariance of the
`singlet soup' the result for an $x$-like or $y$-like Boson
will be precisely the same - we are choosing the $z$-like
component because the ambiguous states in this case are again the ones
with N\'eel order along the loop).
Then, a necessary condition for the scalar product to be
different from zero is that there is a single loop $u_0$
in the resulting loop covering $a+b$ which passes
through both bonds $m$ and $n$. The reason is that the time-reversal
parity of the triplet is negative whereas that of
the singlet is positive. A necessary condition for
a loop $u$ to give a nonvanishing overlap is that the total
time-reversal parities `along the loop' are equal for both states
$|\psi_a\rangle$ and $|\psi_b\rangle$.
This, however, is only possible if the triplets in $|\psi_a\rangle$
and $|\psi_b\rangle$ are within the same loop. Each loop in
$a+b$ therefore must contain either no triplet or both of them.\\
We can now split up the entire overlap integral 
$\langle \psi_a | \psi_{b} \rangle$ into components
which differ by the length and topology of the loop $u_0$
which passes through both triplets. The absolute numerical value 
of the overlap from this particular loop is identical to the case of
pure singlet covering. The only change may be an extra minus sign, which
originates because the singlets do have an orientation, whereas the
triplets do not. We thus can rewrite the overlap as
\begin{eqnarray}
\langle \psi_a | \psi_{b} \rangle
&=&
\sum_{u_o}\;  2^{1-L(u_0)} \; (-1)^{\sigma(u_0)}\; \chi(u_0),
\nonumber \\
\chi(u) &=&  
\frac{ 2^{-(N-L(u))}}{n_1} 
\sum_{a,b} 2^{P(a+b)-1} \Delta_{a+b,u}\;.
\label{over-real}
\end{eqnarray}
Here $\Delta_{c,u}$ is $1$ if the loop covering $c$ contains the
loop $u$ and zero otherwise. We also note that $\chi(1)$$=$$1$,
which fixes the normalization factor $n_1$.
With the exception of the $\chi(u)$
all parts in (\ref{over-real}) can be computed analytically.
$\chi(u)$ may be viewed as the norm of a
nearest-neighbor RVB state which covers only the exterior of the
loop $u$, divided by the norm $n_1$ of the state which covers the
exterior of a single bond.
If we assume that the norm increases exponentially
with the number of sites in the system,
$n \approx e^{\alpha N}$, with $\alpha >0$,
one would estimate that $\chi(u) \approx e^{-\alpha (L(u)-1)}$. 
This suggests that
$\chi(u)$ is a quite rapidly decreasing function of $L(u)$.
In the present work numerical values
for $\chi(u)$ were obtained
by exact calculations on finite clusters
(see the Appendix). The $\chi(u)$ thereby
indeed turned out to decay
rapidly with $L(u)$, so only contributions with 
$L(u)$$\leq$$2$ were kept in the present calculation.
It should be noted
that the computation of the $\chi(u)$
is no fundamental obstacle to the present scheme:
it may well be possible to obtain
essentially exact values for these parameters by
using Monte-Carlo techniques on large lattices.
Figure \ref{fig4} then shows the 
\noindent 
pairs of bonds 
$(i,\alpha)$ and $(j,\beta)$ which can be connected by loops of length
$2$ and $3$ as well as the corresponding signs
$(-1)^{\sigma(u_0)}$. In this way we find
the overlap matrix
\[
N(\bbox{k}) = \sum_{L=1}^{\infty}
\frac{\chi(L)}{2^{L-1}} \gamma_L(\bbox{k})
\]
with $\gamma_1(\bbox{k})$$=$$1$ and
\begin{equation}
\gamma_2(\bbox{k}) =
\left(
\begin{array}{c c}
 0 & 4\sin(\frac{k_x}{2}) \sin(\frac{k_y}{2})\\
4\sin(\frac{k_x}{2}) \sin(\frac{k_y}{2}) & 0
\end{array} \right).
\label{gamma}
\end{equation}
We proceed to a calculation of the matrix elements of the
Hamiltonian. We first recall that for the
nearest neighbor RVB state the expectation value of $H$
between two dimer coverings $|\psi_a\rangle$ and $|\psi_b\rangle$
can be decomposed into contributions from each individual loop
in the loop covering $a+b$\cite{Sutherland}:
\begin{eqnarray*}
\langle \psi_a |H| \psi_{b} \rangle &=& [ \sum_{u \in a+b} E(u) \;]
\;2^{P(a+b)-N},
\nonumber \\
E(u) &=& \epsilon_s \; [ (2 -\delta_{L(u),1} \;)
\;L(u) + n_b(u)\;],
\end{eqnarray*}
where $n_b(u)$ is the number of nearest-neighbor bonds in $u$ 
which bridge the loop (see Figure \ref{fig5}),
and $\epsilon_s$$=$$-3J/4$ 
the exchange energy/singlet.
$E(u)/\epsilon_s$ is  the number of nearest-neighbor
bonds that can be formed from the sites covered by $u$\cite{Sutherland}.
This formula implies that there is no 
\noindent 
contribution from bonds connecting {\em different loops}. The reason is 
that the exchange along a bond connecting different singlets
can only lead to the pair creation of two triplets,
see the first equation (\ref{st}). In order
to maintain time reversal symmetry along
each loop, it is then necessary that both triplets
belong to the same loop - which is not possible if the
bond in question connects different loops.\\
Let us now again assume that the bond $m$ in $|\psi_a\rangle$ and the bond
$n$ in $|\psi_b\rangle$ are occupied by a triplet, and consider the
`connected matrix element' of $H$ between the two resulting states:
$\langle\psi_b |H|\psi_a \rangle 
-E_0 \langle \psi_b|\psi_a \rangle$.
First, let us assume that we act with the exchange along 
a bond connecting $m$ and a neighboring bond $m'$;
the triplet can either propagate from $m$ to $m'$,
or decay into two triplets on both $m$ and $m'$
(see the second Equation (\ref{st})).
Neglecting the second
possibility we obtain a nonvanishing contribution
to the matrix element of $H$ only if there is a single
loop $u_0\in a+b$ which covers both, $m'$ and $n$.
Alternatively, if we act on a bond which connects $n$ and
a neighboring bond $n'$, the triplet jumps from $n$ to $n'$
and we get a nonvanishing contribution only
if one single loop $u_0\in a+b$ covers $n'$ and $m$.
If, on the other hand, we act with the exchange along a bond which
does not touch either of the triplet bonds $m$ or $n$,
both triplets will stay where they are and we get a nonvanishing
matrix element only if both, $m$ and $n$, are covered by a single
loop $u_0\in a+b$. The same holds true if we act with the exchange
along the bonds $m$ and $n$ themselves.\\
We first consider the case that $m$ and $n$ belong to two different loops,
$u_0$ and $u_0'$.
In the simplest case both `loops' consist only of a single
bond, i.e. $u_0$ consists of the single
bond $m$ and $u_0'$ only of $n$. Since the two triplets
belong to different loops, the overlap
$\langle \psi_b|\psi_a\rangle$ is zero.
Moreover, the matrix elements of the exchange along
any bond which does not connect $m$
and $n$ vanishes - the calculation thus becomes very easy.
The matrix element for the triplet hopping from $m$ to
$n$ is $\pm J/4$, where the signs for different
relative positions of the two bonds are shown in Figure \ref{fig6}.
To `embed' the hopping process
\noindent 
into the singlet background,
we need to renormalize this matrix element by
\begin{equation}
\eta_{nm} =
\frac{2^{-N}}{n_1}
\sum_{a,b} 2^{P(a+b)} \Delta_{a+b,m} \Delta_{a+b,n},
\label{etadef}
\end{equation}
which is again estimated from cluster calculations.
The first contribution to the Hamilton matrix then is:
\begin{equation}
\epsilon^{(1)}(\bbox{k})= 
\frac{J \eta}{4} t_{1}(\bbox{k}),
\label{eps1}
\end{equation}
where the elements of the $2\times 2$ matrix $t_1(\bbox{k})$ are:
\begin{eqnarray}
t_{1,xx}(\bbox{k}) &= &
4\cos(k_y) -2 \cos(2k_x) - 4 \cos(k_x) \cos(k_y),
\nonumber \\
t_{1,yy}(\bbox{k}) &=&
4\cos(k_x)- 2 \cos(2k_y) - 4 \cos(k_x) \cos(k_y),
\nonumber \\
t_{1,xy}(\bbox{k}) &=&
4 \sin(\frac{3k_x}{2})\sin(\frac{k_y}{2})
+ 4 \sin(\frac{3k_y}{2})\sin(\frac{k_x}{2}).
\label{kin1}
\end{eqnarray}
To keep things simple we have moreover replaced the different
$\eta_{nm}$ by the average value $\eta$.\\
Next, we consider the case that one of the loops,
e.g. $u_0'$, has a length $\ge 2$.
In other words, we a considering a process like the one shown in
Figure \ref{fig7}: the triplet jumps from bond $m$ to bond $m'$,
and the triplet on $m'$ 
\noindent 
overlaps with the triplet
on bond $n$ along a loop (in this case of length $2$).
There is also an analogous process, where the triplet jumps
from $n$ to $m'$ and $m'$ and $m$ overlap by a loop.
The respective matrix elements can be factorized
into the matrix element for the hopping of the triplet
times the overlap along the loop.
This means that the matrix element can be written as
$\frac{\eta_3 J}{8} t_2(m,n)$ where
\[
t_2(m,n) = \sum_{m'} [ t_1(m,m') \gamma_2(m',n) +
t_1(n,m') \gamma_2(m',m)];
\]
here $t_1(m,m')$ and $\gamma_2(m',n)$
are the real-space versions of the matrices (\ref{kin1}) and
(\ref{gamma}). Fourier transformation gives 
\begin{equation}
\epsilon^{(2)}(\bbox{k}) = \frac{\eta_3 J}{8}
( t_1(\bbox{k}) \gamma_2(\bbox{k}) +\gamma_2(\bbox{k})t_1(\bbox{k}))
\label{eps2}
\end{equation} 
where it has to be kept in mind that $t_1(\bbox{k})$ and
$\gamma_2(\bbox{k})$ are {\em symmetric}
$2\times 2$ matrices. The `embedding factor' is
\begin{equation}
\eta_{3} = 
\frac{2^{-N}}{n_1}
\sum_{a,b} 2^{P(a+b)} \Delta_{a+b,m}\Delta_{a+b,n'} \Delta_{a+b,n} 
\label{eta3def}
\end{equation}
where $(m,n',n)$ are like in Figure \ref{fig7}.
Actually there are two inequivalent relative orientations
of a single bond $m$ and an $L=2$ loop - the two respective values
of $\eta_3$ do not differ strongly and for simplicity we
use the average of the two values for the two configurations.
Processes involving even longer loops could be treated in an analogous 
way, but we neglect these.\\
We proceed to the second case, i.e. we assume that $m$ and $n$ are
covered by a single loop $u_0$.
As was the case for the overlap integrals, the matrix
elements then can be split up
into contributions differing by the loop $u_0$ which
covers both triplet bonds. Once this loop
is fixed, we can divide all bonds in the
plane into three distinct classes.
First, bonds belonging to any loop other than $u_0$ are not affected at all
by the presence of the triplets and give the same contribution
as in the pure singlet state. This becomes
\begin{eqnarray}
2^{1-L(u_0)} &(&-1)^{\sigma(u_0)}
\frac{ 2^{-(N-L(u_0))}}{n_1} 
\nonumber \\
&\;&\;\;\;\;
\sum_{a,b}  [ 
\sum_{\scriptsize \begin{array}{c}
 u \in a+b\\
u \neq u_0
\end{array} } E(u) \;] 2^{P(a+b)-1} \Delta_{a+b,u_0}.
\label{bulk1}
\end{eqnarray}
This is an energy of order of the system
size $N$; in the end it must be canceled, up to terms of order $N^0$, by
a corresponding contribution in $-E_0 \langle \psi_a | \psi_{b} \rangle$.
This cancellation is the analogue of the familiar linked-cluster
theorem of many-body physics.\\
Second, we consider those bonds in $u_0$ which are not bridging bonds. 
They will be covered by either a singlet or a triplet
in either $a$ or $b$, whence these bonds together give the
contribution
\begin{eqnarray}
\;[ E(u_0) &+& (2 -\delta_{L(u_0),1} \;)J - \epsilon_s n_b(u_0)]
\nonumber \\
&\;&\;\;\;\;\;\;\;\; 2^{1-L(u_0)} (-1)^{\sigma(u_0)} \chi(u_0).
\end{eqnarray}
This is a `connected' contribution of order $N^0$.\\
This leaves us with the bridging bonds (see
Figure \ref{fig5}), which may give a
nontrivial contribution. However, since the bridging
bonds occur only for $L(u_0) \ge 3$ we neglect 
their contribution altogether.\\
The subtracted contribution,
$-E_0 \langle \psi_a | \psi_{b} \rangle$, may be rewritten as
\begin{eqnarray}
- E_0\;\sum_{u_0} 
2^{1-L(u_0)} &(&-1)^{\sigma(u_0)}
\frac{ 2^{-(N-L(u_0))}}{n_1} 
\nonumber \\
&\;&\;\;\;\;\;\;\;
\sum_{a,b} 2^{P(a+b)-1} \Delta_{a+b,u_0},
\label{bulk2}
\end{eqnarray}
where we have used the expanded form (\ref{over-real}) of
$\chi(u_0)$.
This is again an energy of order $N$, which
cancels the bulk term (\ref{bulk1}) up to terms of order
$N^0$. After some reshuffling (using $\sum L(u)$$=$$N$) we can 
rewrite the contribution to the matrix element as
\begin{equation}
J \langle \psi_a | \psi_b\rangle +
\sum_{u_0} (\epsilon_J(u_0)+ \epsilon_{0}(u_0)\;)
\label{hmat}
\end{equation}
with
\begin{eqnarray}
\epsilon_J(u) &=& J\;(1 -\delta_{L(u_0),1} \;)\;
2^{1-L(u)} (-1)^{\sigma(u)} \chi(u)
\nonumber \\
\epsilon_{0}(u)
&=&  \frac{ (-1)^{\sigma(u)}}{n_1} \sum_{a,b} 
[\;\sum_{u\in a+b}\bar{E}(u)\;]
2^{P(a+b)-N} \Delta_{a+b,u},
\nonumber \\
\bar{E}(u) &=& E(u) - L(u) \frac{E_0}{N}.
\label{mats}
\end{eqnarray}
The first term on the rhs of (\ref{hmat}) is the `bare'
on-site energy of the triplet. It  is always proportional to the
overlap integral, so upon switching to the
effective Bosons this term becomes a
$\bbox{k}$-independent constant shift. 
The quantity $\epsilon_0(u)$ may be thought of as
describing a `loss of resonance energy'.
It is the difference of the two contributions
(\ref{bulk1}) and (\ref{bulk2}) and
occurs because the loop $u_0$ is fixed, whence the 
area covered by this loop is not available for resonating
between different singlet coverings. For example
we have
\begin{equation}
J + \epsilon_0(1)
= \frac{\langle \psi_a | H |  \psi_a \rangle}
{\psi_a |  \psi_a\rangle} - E_0,
\label{renorm}
\end{equation}
i.e. this quantity is an additive
renormalization of the energy of formation for a single triplet
due to its being `embedded into the singlet soup'. 
Numerical evaluation in a cluster shows that this additive correction
is quite large -
for one triplet in a pure singlet background
we find $\epsilon_0(1)\approx 1.2J$. While this is surprising at first sight
it should be noted that a similar large value
($\approx 0.8J$) was previously found in spin ladders\cite{lad}.
A fixed triplet obviously leads to a quite substantial loss
of resonance energy.
The numerical values of $\epsilon_0(u)$ and $E_0$ were again obtained by
cluster calculations (Appendix). 
Introducing $\tilde{\epsilon}_0(u)$$=$$
\epsilon_0(u) + J \chi(u)$ we can write down
the third part of the Hamilton matrix:
\begin{equation}
\tilde{\epsilon}^{(3)}(\bbox{k}) =
\epsilon_0(1) + \sum_{L\ge 2}
\frac{\tilde{\epsilon}_0(L)}{2^{L-1}} \gamma_L(\bbox{k})
\label{eps3}
\end{equation}
We can now add up the three contributions,
(\ref{eps1}), (\ref{eps2}), and (\ref{eps3})
to obtain the total `connected' Hamilton matrix $\tilde{\epsilon}(\bbox{k})$.
This is still expressed in terms of the non-orthogonal orbitals
$\tilde{\tau}_{\bbox{k},\alpha}^\dagger$.
What remains to be done therefore is to transform the Hamilton matrix to the
orthogonal orbitals (\ref{ortho}). To that end
we take matrix elements of the form
$\langle \nu, \bbox{k}| \tilde{\epsilon}(\bbox{k})| \nu', \bbox{k}\rangle$.
Introducing the $2\times 2$ transformation matrix
\begin{equation}
T = \left(
\begin{array}{c c}
\frac{1}{\sqrt{\lambda_1}} \bbox{e}_1, \frac{1}{\sqrt{\lambda_2}} \bbox{e}_2
\end{array} \right),
\end{equation}
the transformed Hamiltonian then can be expressed as
\[
\epsilon(\bbox{k}) = T^T \tilde{\epsilon}(\bbox{k}) T.
\]
\section{Pair creation amplitude}
\begin{figure}
\epsfxsize=8.0cm
\vspace{-0.5cm}
\hspace{-0.5cm}\epsfig{file=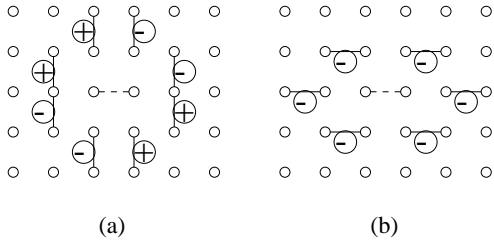,width=8.0cm,angle=0.0}
\vspace{0.0cm}
\narrowtext
\caption[]{
The sign $(-1)^{\sigma(m,n)}$ for all different pairs
of bonds on which pair creation is possible. The bond
$m$ (dashed) is kept fixed, the bond $n$ is
labeled by $(-1)^{\sigma(m,n)}$.}
\label{fig8} 
\end{figure}

\begin{figure}
\epsfxsize=4.0cm
\vspace{-0.0cm}
\hspace{1.0cm}\epsfig{file=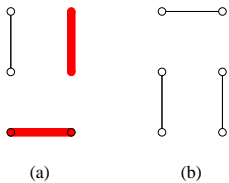,width=4.0cm,angle=0.0}
\vspace{0.5cm}
\narrowtext
\caption[]{A state with two triplets (a) can have a nonvanishing
overlap with a state consisting exclusively of two
singlets (b).}
\label{fig9} 
\end{figure}

In this section we proceed to a discussion of the pair creation amplitude
$\Delta_{\bbox{k}}$ in equation
(\ref{heff}). As discussed in section II, by starting from an
arbitrary singlet covering of the plane and acting with the
Hamiltonian along a bond connecting two different singlets,
we {\em only} create a state where both singlets are replaced
by triplets, see the first of Eqs. (\ref{st}).
The situation where the two singlets in question are `parallel' to
each other, i.e. that the $4$ sites belonging to the
$2$ singlets form a square with edge $1$, requires special attention.
As discussed above we have the identity
\begin{equation}
s_{i,j}^\dagger s_{i',j'}^\dagger |0\rangle 
- 2 s_{i,i'}^\dagger s_{j,j'}^\dagger |0\rangle 
= \bbox{t}_{i,j}^\dagger \cdot \bbox{t}_{i',j'}^\dagger |0\rangle,
\label{combo}
\end{equation}
i.e. the state with two parallel triplets can be expressed
as a linear combination of the two perpendicular combinations of
parallel singlet states.
In other words: this state is 
already exhausted by the singlet background, and
consequently must be omitted from our reduced Hilbert space.
The triplets thus have to obey the additional constraint
of never being parallel to each other - one implication
is that we must set the respective pair creation matrix element
to zero. \\
To compute the numerical value of the matrix element,
let us consider the action of $H$ on the RVB state (\ref{RVB}).
We have
\begin{eqnarray}
H |RVB\rangle &=&
 N \epsilon_s |RVB\rangle
\nonumber \\
&\;&\;\;\;+\frac{J(-1)^{\sigma(m,n)}}{4} \frac{n_1}{\sqrt{n}} 
\sum_{m,n} \sum_{\alpha}'
|\Phi_{m\alpha,n\alpha}\rangle.
\label{hrvb}
\end{eqnarray}
The first term on the rhs originates from processes where the
Hamiltonian has `hit' a bond covered by a singlet, the second
one originates from processes where the exchange has acted along
bonds connecting two singlets on bonds $m$ and $n$ (the prime
on the sum indicates that
only pairs of bonds connected by a bond are summed over).
The modulus of the respective matrix element is $J/4$
(see equation (\ref{st})) and there is an extra sign
which depends on the relative orientation of the bonds $m$ and $n$.
The dependence of this sign on the orientation is shown in Figure
\ref{fig8}. Also, we have approximated the normalization
factor which is included in the definition of
$|\Phi_{m\alpha,n\alpha}\rangle$ by
\begin{equation}
\frac{1}{n(m\alpha,n\alpha)} \approx \frac{1}{\sqrt{n_1}^2}.
\end{equation}
The factor of $1/\sqrt{n}$ remains from the definition of (\ref{RVB}).
To obtain the pair creation matrix element we now
form the overlap between (\ref{hrvb}) and the
state
$\tau_{\bbox{k},\nu}^\dagger \tau_{-\bbox{k},\mu}^\dagger|0\rangle$.
\noindent 
Defining the $2\times 2$ matrix
\begin{eqnarray}
\tilde{\Delta}(\bbox{k})
&=& \frac{\zeta J}{8} t'_{1,\alpha\beta}(\bbox{k}),
\nonumber \\
\zeta &=& \frac{n_1}{\sqrt{n}} ,\nonumber \\
t'_{1,xx}(\bbox{k}) &= &
-2 \cos(2k_x) - 4 \cos(k_x) \cos(k_y),
\nonumber \\
t'_{1,yy}(\bbox{k}) &= &
-2 \cos(2k_y) - 4 \cos(k_x) \cos(k_y),
\nonumber \\
t'_{1,xy}(\bbox{k}) &= &
4 \sin(\frac{3k_x}{2})\sin(\frac{k_y}{2})
+ 4 \sin(\frac{3k_y}{2})\sin(\frac{k_x}{2}),
\label{pair}
\end{eqnarray}
and bearing in mind that $\langle \bar{\tau}_{\bbox{k},\alpha}^{}|
\bar{\tau}_{\bbox{k},\alpha'}^\dagger \rangle = N_{\alpha,\alpha'}
(\bbox{k})$,
we find for the pair creation matrix:
\[
\Delta(\bbox{k}) = T^T N(\bbox{k})\tilde{\Delta}(\bbox{k})N^T(\bbox{k}) T.
\]
This completes the derivation of the single-particle
terms of the triplet Hamiltonian. Before we proceed,
let us briefly return to the problem of non-orthogonality.
Strictly speaking, the state with two triplets is not 
orthogonal to the singlet background either.
The reason is that if one draws a loop passing through both
triplets, the time reversal parities of the two triplets
cancel, and the state has a nonvanishing overlap with a state where the
loop is covered only by singlets (see Figure \ref{fig9}).
\noindent 
However, this overlap is rather small:
in the case shown in Figure (\ref{fig9})
it is for example $-\sqrt{3}\chi(3)/8$,
and obviously this is the most unfavorable case.
For other relative orientations of the two triplets
the overlap can be only achieved by a loop of length $8$, whence these
overlaps are $\propto \chi(4)\ll 1$. The non-orthogonalities thus
are much more benign that those for the spinon-states, whence
neglecting them altogether (as we will
do henceforth) is probably quite justified.
\section{Extrapolation to Finite Triplet Density}
In the preceding sections we have computed the
various overlap integrals and matrix elements
for triplet propagation, pair creation and interaction.
In all of these cases the matrix element could be factorized into a
contribution from a `local' transition between
different singlet/triplet coverings along a single or two
neighboring loops,
and a factor which describes the embedding of these active loops into
the singlet background. Thereby we have always given expressions
for these `embedding factors'
which are valid in the limit of vanishing triplet concentration,
i.e. we have computed them as they would be
for a pure singlet covering of the system. Clearly, this is
inappropriate for the real system, where quantum fluctuations
have admixed a finite density of triplet
Bosons. In the following we want to discuss
how we have to modify our theory to take the effect of a finite
triplet concentration into account.  It should be stressed
from the very beginning that this is quite obviously a very complex
problem and we will be forced to apply some relatively
crude approximations.\\
If we want to derive single-particle like matrix elements
for finite triplet concentration we should consider
overlap integrals or matrix elements of $H$ of the type
\begin{equation}
\langle \Psi_{j_1\alpha_1, i_2\alpha_2,\dots i_m\alpha_m}
|\Psi_{i_1\alpha_1, i_2\alpha_2,\dots i_m\alpha_m}\rangle,
\end{equation}
i.e. $m-1$ triplets stay unchanged, and only a single one
(which without loss of generality can be taken to be
the first one) changes its position (but maintains its
spin). In other words, we should calculate the embedding factors
for singlet coverings containing a certain number
of `inert' triplets. Thereby we actually have to make the
major {\em assumption} that the matrix element does not depend significantly 
on the positions of these inert triplets - otherwise, the very idea of a
single particle-like propagation of the triplets would be
invalid. One might then expect that approximate values can be obtained
by distributing the $m$ inert triplets in all possible ways
(we call the number of possible distributions D(m))
over the system, and taking the average of the respective
matrix elements computed for all D(m) possible distributions.
In this way, the embedding factors acquire a dependence
on the density of triplets.\\
The numerical calculation in a finite cluster then proceeds as follows:
we choose $m$ bonds $i_1, i_2\dots i_m$, which obey the
various constraints on the relative positions of triplets,
and evaluate the ground state norm and energy according to
\begin{eqnarray}
n &=& \sum_{a,b} 2^{P(a+b)-N}\prod_{\nu=1}^{m} \Delta_{a+b,i_\nu},
\nonumber \\
E_0 &=& \sum_{a,b} 
[\;\sum_{u\in a+b}
\epsilon(u)\;]\;2^{P(a+b)-N}\prod_{\nu=1}^{m} \Delta_{a+b,i_\nu}.
\label{hugo}
\end{eqnarray}
The calculation of the various embedding factors then
proceeds in an entirely analogous fashion i.e.
in (\ref{over-real}), (\ref{etadef}), and (\ref{mats})
 we replace 
\begin{eqnarray}
\sum_{a,b} \dots \rightarrow \sum_{a,b}
\left( \prod_{\nu=1}^{m} \Delta_{a+b,i_\nu} \right) \dots
\end{eqnarray}
In this way we obtain all embedding factors
for fixed distribution of inert triplets, and the
value for triplet concentration $m/N$ is obtained by 
averaging over all allowed distributions of the bonds
$i_1, i_2\dots i_m$. In practice this calculation
requires quite a substantial
numerical effort so this was performed only for the
$4\times 4$ cluster.
\begin{figure}
\epsfxsize=4.0cm
\vspace{-1.0cm}
\hspace{1.0cm}\epsfig{file=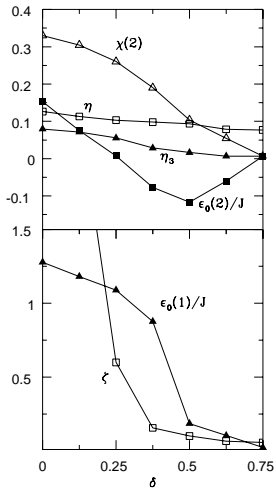,width=8.0cm,angle=0.0}
\vspace{-0.5cm}
\narrowtext
\caption[]{Dependence of the various parameters
on the triplet density $\delta$ as obtained
from numerical calculations on the $4\times 4$ cluster.}
\label{fig10} 
\end{figure}
\noindent 
One might expect, however, that the finite size effects are in fact smaller
for the more relevant higher triplet concentrations:
the main effect of the fixed triplets obviously is to reduce the
importance of long loops. Since the bonds occupied by the static triplets 
must be identical in the bra and ket state, loops of length
$L(u)\ge2$ which pass through these bonds are impossible.
When the density of triplets gets appreciable, the probability to
find enough space for forming longer loops becomes smaller and
smaller. Since longer loops may cause problems in the small
clusters, a suppression of these may therefore even reduce finite-size
effects. Moreover, the suppression of longer loops
is actually beneficial for our entire
approach: it tends to eliminate the problems 
with the nonorthogonalities in the singlet soup and makes
the truncation of the length $L(u_0)$ a better approximation.\\
As an illustration Figure \ref{fig10} shows the dependence of
$\epsilon_0(1)$, i.e. the additive renormalization of the
energy of formation of a triplet (see (\ref{renorm}), evaluated in the 
$4\times 4$ cluster.
The concentration dependence is as expected:
for low triplet concentration $\epsilon_0(1)$ is large
because an added triplets blocks many long loops along which
resonance could have occurred. As the triplet concentration gets higher,
these longer loops a blocked anyway, so adding a further triplet does
not increase the energy too much any more. In the high-density limit
the additive renormalization approaches zero, as expected. The Figure
shows, however, that the concentration dependence is quite significant,
i.e. this effect should not be neglected.\\
In addition to reducing the importance of longer loops,
for a finite density of triplets we have to take care of the
excluded volume constraint which the bond Bosons have to
obey. Placing a triplet on one given bond $m$ blocks a total
of $9$ other bonds, on which no more triplets can be placed (see Figure
\ref{fig11}).
\begin{figure}
\epsfxsize=4.0cm
\vspace{-0.0cm}
\hspace{1.0cm}\epsfig{file=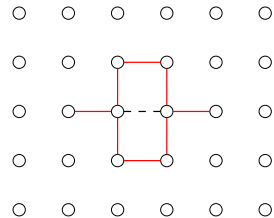,width=4.0cm,angle=0.0}
\vspace{0.5cm}
\narrowtext
\caption[]{Placing a triplet on the central (dashed) bond blocks the
$9$ indicated bonds, on which no more triplet can be placed.}
\label{fig11} 
\end{figure}
\noindent 
Of these, $7$ bonds are blocked because they share a site
with $m$, the remaining $2$ are blocked because they are `parallel'
to $m$ and the state with two parallel triplets is actually
a linear combination of singlets (see (\ref{combo})). 
In order to take care of this
{\em blocking effect}, we resort to a Gutzwiller-type approximation.
It has been shown by  Ogawa {\em et al.}\cite{ogawa} that the essence
of the Gutzwiller approximation is the neglect of the
difference in phase between states where the particles
in question are distributed in different ways over the lattice.
With this approximation, any real space
distribution of $m$ Bosons contributes the same 
number (which without loss of generality
can be taken to be $1$) to the norm of any state with $m$ Bosons.
The total norm of any such state then becomes simply the number of
possible ways to distribute $m$ Bosons over the plane, i.e. $D(m)$.
In the present case, the situation is somewhat more
complicated, because we still have to take into account
the fact that due to the loop problem the norm of a state
with $m$ fixed triplets is not $1$ (unlike the case
if the triplets were just ordinary Bosons).
We thus approximate the normalization factor
 of any state with $m$ triplets by
\begin{equation}
n(m) = \frac{1}{\sqrt{D(m)\bar{n_m}}},
\end{equation}
where $\bar{n_m}$ is the value obtained by
averaging (\ref{hugo}) over all allowed triplet configurations
$i_1, i_2\dots i_m$. Clearly, this replacement
will lead to a renormalization
of the various embedding factors. To evaluate
these, we usually have to
consider a certain area ${\cal A}$, covered by
one or two loops, in which the actual
overlap, hopping or pair creation process takes place.
We then determine (again by simulation on the $4\times 4$ cluster) the
number of ways  to distribute $m$ triplets
over the exterior of the area  ${\cal A}$. Thereby we request
that putting a triplet anywhere within  ${\cal A}$ always gives
an allowed configuration of $m+1$ triplets.
We call the number of allowed triplet configurations
$D(m,{\cal A})$. Defining the various areas ${\cal A}_i$
as 
\begin{figure}
\epsfxsize=4.0cm
\vspace{-0.0cm}
\hspace{1.0cm}\epsfig{file=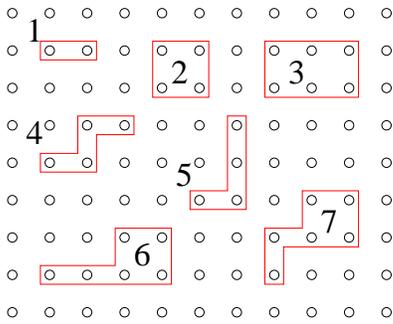,width=6.0cm,angle=0.0}
\vspace{0.5cm}
\narrowtext
\caption[]{Areas on which no triplets are allowed in some
processes.}
\label{fig12} 
\end{figure}
\noindent 
in figure \ref{fig12}, we then have to renormalize the various parameters
as follows:
\begin{eqnarray}
n_1 \rightarrow \frac{n_1}{\sqrt{D(m,{\cal A}_1)}},
\nonumber \\
\chi(2) \rightarrow \chi(2) \frac{D(m,{\cal A}_2)}{D(m,{\cal A}_1)},
\nonumber \\
\epsilon_0(2) \rightarrow \epsilon_0(2)
\frac{D(m,{\cal A}_2)}{D(m,{\cal A}_1)},
\nonumber \\
\eta \rightarrow \eta \frac{D(m,{\cal A}_4) +D(m,{\cal A}_5) }
{2D(m,{\cal A}_1)},
\nonumber \\
\eta_{3} \rightarrow \eta_{3} \frac{D(m,{\cal A}_6)+D(m,{\cal A}_7)}
{2D(m,{\cal A}_1)}.
\end{eqnarray}
For the pair creation amplitude we replace
\begin{equation}
\zeta \rightarrow \frac{\zeta}{2}
(\sqrt{ \frac{D(m,{\cal A}_4)}{n}}+
\sqrt{ \frac{D(m,{\cal A}_5)}{n}})
\end{equation}
To summarize this section, Figure \ref{fig10} shows  the values
of the different parameters as estimated by the procedure outlined
above, for all possible triplet concentrations in the $4\times 4$ cluster.
Let us stress again that the approximations leading to
the parameters in Figure \ref{fig10} are probably rather crude - one may 
expect, however that we get roughly correct orders of magnitude and
that the ratios of the different parameters come out approximately
correct.
\section{Spin dynamics}
\begin{figure}
\epsfxsize=5.0cm
\vspace{-0.5cm}
\hspace{1.0cm}\epsfig{file=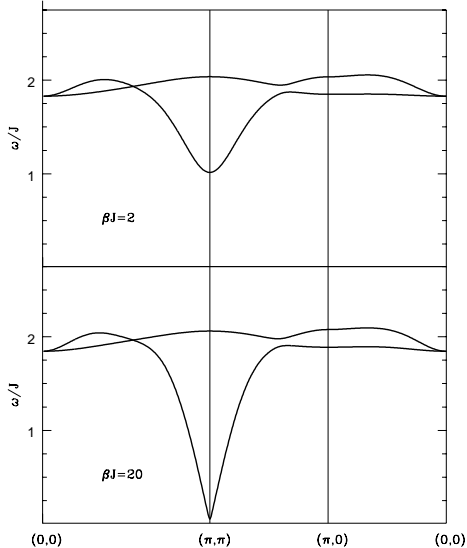,width=8.0cm,angle=0.0}
\vspace{0.0cm}
\narrowtext
\caption[]{Dispersion of the two eigenvalues $\omega_i$, for two
different temperatures.}
\label{fig13} 
\end{figure}

Let us first briefly summarize the discussion so far:
We have shown that the elementary excitations of the
singlet soup correspond to excited dimers, which must be
modeled by bond-Bosons. 
Combining the matrix elements computed in the two preceding sections
we obtain a Hamiltonian of the form
\[
H = \sum_{\bbox{k},\nu,\mu}
\epsilon(\bbox{k})_{\nu,\mu} \bbox{\tau}_{\bbox{k},\nu}^\dagger \cdot
\bbox{\tau}_{\bbox{k},\mu}^{}
+ 
(\Delta(\bbox{k})_{\nu,\mu} \bbox{\tau}_{\bbox{k},\nu}^\dagger \cdot
\bbox{\tau}_{-\bbox{k},\mu}^\dagger + H.c.).
\]
Thereby the matrix elements
$\Delta(\bbox{k})$ and $\epsilon(\bbox{k})$ are functions of the
triplet density.
For given triplet density $\delta_{i}$
these parameters can be computed and
the Hamiltonian, which after the Gutzwiller-type
renormalization of the matrix elements we take to
be a free-Boson Hamiltonian, is solved by Bogoliubov transformation.
Combining the two different $\tau$-operators into a 2-vector
$\bbox{T}$ the ansatz reads
\begin{equation}
\bbox{\Gamma}_{\bbox{k}}^{} = u_{\bbox{k}} \bbox{T}_{\bbox{k}}^{}
+ v_{\bbox{k}} \bbox{T}_{-\bbox{k}}^\dagger,
\label{bogo}
\end{equation}
where the real $2\times2$ matrices $u$ and $v$ have to fulfill
\begin{eqnarray}
u_{\bbox{k}}^{} u^T_{\bbox{k}} - v_{\bbox{k}}^{} v^T_{\bbox{k}} &=& 1,
\nonumber \\
u_{\bbox{k}}^{} v^T_{\bbox{k}} - v_{\bbox{k}}^{}      u^T_{\bbox{k}} &=& 0.
\end{eqnarray}
The inverse transformation of (\ref{bogo}) is therefore
\begin{equation}
\bbox{T}_{\bbox{k}}^{} = u_{\bbox{k}}^T \bbox{\Gamma}_{\bbox{k}}^{}
- v_{\bbox{k}}^T \bbox{\Gamma}_{-\bbox{k}}^\dagger,
\label{bogo_inv}
\end{equation}
The Hamiltonian can be transformed to free particle form:
\begin{eqnarray}
H &=& \sum_{\bbox{k},\nu}
\omega_\nu(\bbox{k}) \bbox{\Gamma}_{\bbox{k},\nu}^\dagger
\bbox{\Gamma}_{\bbox{k},\nu}^{} +
\nonumber \\
&+&
3 \sum_{\bbox{k}} Tr[ v_{\bbox{k}}\epsilon(\bbox{k}) v^T_{\bbox{k}}  
- v_{\bbox{k}} \Delta(\bbox{k}) u^T_{\bbox{k}} 
- u_{\bbox{k}} \Delta(\bbox{k}) v^T_{\bbox{k}}],
\end{eqnarray}
provided the transformation matrices obey
\begin{eqnarray}
\epsilon(\bbox{k}) u^T_{\bbox{k}} - 2\Delta(\bbox{k}) v^T_{\bbox{k}} 
&=&  u^T_{\bbox{k}} \omega(\bbox{k}) ,
\nonumber \\
2 \Delta(\bbox{k}) u^T_{\bbox{k}} - \epsilon(\bbox{k}) v^T_{\bbox{k}} 
&=&  v^T_{\bbox{k}} \omega(\bbox{k}).
\label{eigenvalue}
\end{eqnarray}
Here we have introduced
the $2\times 2$ matrix $\omega(\bbox{k})=diag(\omega_1(\bbox{k}),
\omega_2(\bbox{k})$.
The triplet density is
\begin{equation}
\delta_{f} =
\frac{3}{N}\sum_{\bbox{k}}
Tr[ u_{\bbox{k}} u^T_{\bbox{k}} f(\omega) + v_{\bbox{k}} 
v^T_{\bbox{k}} (1+f(\omega)) ],
\label{den}
\end{equation}
where $f$ is the Bose function.
The requirement $\delta_{i}=\delta_{f}$
then provides a self-consistency condition for the density.\\
Numerical evaluation shows that there is a minimum triplet concentration
$\delta_0$. 
For $\delta < \delta_0$, the parameters are such that
in a certain area around $\bbox{k}=(\pi,\pi)$ the
eigenvalue problem (\ref{eigenvalue}) does not have real eigenvalues.
For $\delta_0$ the minimum of the triplet dispersion, which occurs
at $(\pi,\pi)$ is precisely zero. 
With the coarse mesh of concentrations possible in the $16$-site
cluster, we find
$0.25 < \delta_0 < 0.375$. We therefore linearize all 
matrix elements $x$, i.e. $x(\delta)= a_x + b_x(\delta-0.25)$,
using the values at $\delta=0.25$ and $\delta=0.375$ to determine
$a_x$ and $b_x$. This gives $\delta_0=0.335$.\\
With increasing temperature,
the self-consistent value of $\delta$ increases slowly, approaching
$\delta_0$ for $T\rightarrow 0$. 
Figure \ref{fig13} shows the dispersion relation for the respective
self-consistently determined  triplet density at different temperatures.
Since we have two degrees of freedom/site (the triplet in $x$
and $y$ direction) we obtain two bands. While one of these bands
is practically dispersionless, the dispersive one
resembles results obtained for other spin 
\noindent
liquids, such as the two-leg ladder\cite{Gopalan,lad} or the
bi-layer Heisenberg antiferromagnet\cite{Kotov}: the dispersion
starts at relatively high energy
at $\bbox{k}=(0,0)$, takes a shallow maximum near the
antiferromagnetic zone boundary and, as stated above, 
takes a more or less pronounced
minimum at $\bbox{Q}=(\pi,\pi)$. For
low temperatures the bandwidth is $\approx 2J$,
which is quite close to the value for antiferromagnetic
spin waves. The case of zero temperature requires special
attention: it is obvious from Figure  \ref{fig13}
that the gap at $\bbox{Q}$ approaches zero for
$T \rightarrow 0$. Assuming that the gap vanishes
at zero temperature, we may then assume that this momentum
becomes macroscopically occupied by a triplet density
$\delta_{\bbox{Q}}$. To determine this condensed fraction we note
that the limiting density
$\delta_{0}$ is defined such that the
gap at $\bbox{Q}$ is exactly zero with the
parameters calculated for this density. Then, for $\delta_i=\delta_0$
we evaluate the density $\bar{\delta}_f$
of uncondensed triplets (i.e. we exclude $\bbox{k}=\bbox{Q}$
from the sum in (\ref{den}))
and finally determine the condensate density from
\begin{equation}
\delta_{\bbox{Q}} = \delta_0 - \bar{\delta}_f.
\label{conden}
\end{equation}
This is entirely analogous to the treatment of 
Hirsch and Tang\cite{hirschtang}
of the condensation of Schwinger-Bosons in the mean-field theory
of Arovas and Auerbach\cite{arovasauerbach}. 
As will be seen in a moment, just as
for the Schwinger Bosons the condensation of the triplets
corresponds to antiferromagnetic ordering. 
The ground state energy then is
\begin{equation}
E_{tot} = E_0(\delta) -
3 \sum_{\bbox{k}} Tr[ v_{\bbox{k}} v^T_{\bbox{k}}  \omega(\bbox{k})]
\end{equation}
where $E_{0}(\delta)$ is the energy of the singlet background
for triplet density $\delta$. This is calculated from
 (\ref{hugo}). We obtain a value of
$-0.3426J$/bond, which is quite close to the true ground state
energy of the 2D Heisenberg antiferromagnet\cite{manousakis}. 
An interesting figure is the lowering of the energy
as compared to the original RVB-vacuum (\ref{RVB}).
In the $4\times 4$ cluster (which has been used to compute
all matrix elements) the expectation value of the pure
nearest-neighbor RVB state is $-0.334318J$/bond\cite{Kohmoto},
so that the admixture of the triplets lowers the energy only by 
a tiny $0.008J$/bond. In view
of the strong approximations we were forced to make
this result probably has little quantitative significance - it shows
quite clearly, however, that the energy of the RVB-vacuum
is lowered only by a very small energy by the triplet
fluctuations. This is what must come out because
in the thermodynamic limit the energy of the RVB state is
$-0.302J$/bond\cite{LiangDoucotAnderson}, 
compared to the exact ground state energy of
approximately $0.334J$/bond\cite{manousakis}.\\
We proceed to a discussion of the spin correlation function.
Since the bond-Bosons are actually
objects which extend over more than one unit-cell, they have
something like a structure factor which depends on the
type of operator by which they are probed. 
Let us consider a single bond, $(i,j)$,
and introduce even and odd combinations of spin operators on this bond:
\[
S_\pm^z =  S_i^z \pm S_j^z.
\]
Due to their opposite parity under exchange of
$i$ and $j$ these two operators have an entirely
different effect on the $4$ possible spin states of the bond:
$S_+^z$ annihilates the singlet and $t_z^\dagger$ but converts
$t_x^\dagger \rightarrow i t_y$ and
$t_y^\dagger \rightarrow -i t_x$, whence
$S_+^z$$=$$ i \bar{\bbox{\tau}}^\dagger \times \bar{\bbox{\tau}}$.
$S_-^z$ on the other hand converts the singlet into
$t_z^\dagger$ and vice versa, but annihilates both
$t_x^\dagger$ and $S_+^z$, whence
$S_+^z$$=$$\bar{\bbox{\tau}}^\dagger + \bar{\bbox{\tau}}$.
If we still restrict ourselves to only a single bond 
$n$ we have to `translate'
\begin{eqnarray}
\sum_j e^{i \bbox{q} \cdot \bbox{R}_j} \bbox{S}_j &=&
i \sum_n e^{i \bbox{q} \cdot \bbox{R}_n} 
\; [ \cos \left( \frac{ \bbox{e}_n \cdot \bbox{q}}{2} \right)
 \bar{\bbox{\tau}}_n^\dagger \times \bar{\bbox{\tau}}_n
\nonumber \\
&\;&\;\;\;\;\;\;\;\;\;\;
- \sin \left( \frac{ \bbox{e}_n \cdot \bbox{q}}{2} \right)
(\bar{\bbox{\tau}}_n^\dagger + \bar{\bbox{\tau}}_n) ],
\label{corr1}
\end{eqnarray}
where $\bbox{q}_n$ is the unit vector along the bond $n$
and $\bbox{R}_n$ denotes its center of gravity.
Let us next discuss how these matrix elements are
modified due to the embedding into the singlet background.
We act with the
operator $\bbox{S}(\bbox{q})=(1/\sqrt{2N} )
\sum_j e^{i \bbox{q} \cdot \bbox{R}_j} \bbox{S}_j$
on some given state. The first term simply changes the
$z$-spin and momentum of a triplet. It does not affect the
number of triplets, so we need no further
renormalizations.
For the second term, let us start out from a state with one
triplet, $|j,\alpha\rangle$, and assume that the triplet is
annihilated, i.e. converted into a singlet.
This leaves us with the state 
\[
|\psi_j\rangle = \frac{1}{\sqrt{n_1}}\frac{S^{N-1}}{(N-1)!} 
s_{j,j+\hat{\alpha}}^\dagger |0\rangle.
\]
The matrix element with the pure RVB state then is
\[
\langle RVB |\psi_j\rangle = \frac{1}{\sqrt{n n_1}} \frac{n}{4}
= \frac{1}{4} \sqrt{\frac{n}{n_1}},
\]
because it is easy to see that a state with one fixed singlet has an overlap
of $n/4$ with the RVB state.
If we want to extrapolate this to finite triplet density
we again have to use the values of $n$ and $n_1$ computed with
$m$ fixed triplets, and renormalize the matrix element by
\[
\kappa= \sqrt{ \frac{D(m,{\cal A}_1)}{D(m)}}.
\]
We thus find
\begin{eqnarray}
\bbox{S}(\bbox{q})
&=&\sum_{\alpha=x,y}\; [\cos(\frac{q_\alpha}{2})
\frac{1}{\sqrt{2N}}\sum_{\bbox{k}}
\bar{\bbox{\tau}}_{\bbox{q}+\bbox{k},\alpha}^\dagger \times 
\bar{\bbox{\tau}}_{\bbox{k},\alpha}^{} 
\nonumber \\
&\;&\;\;\;\;\;\;\;\;\;\;
- \frac{\kappa}{4}\sqrt{\frac{n}{n_1}} \sin(\frac{q_\alpha}{2})
(\bar{\bbox{\tau}}_{\bbox{q},\alpha}^\dagger + 
\bar{\bbox{\tau}}_{\bbox{q},\alpha}^{}) ].
\end{eqnarray}
By using the inverse transformation (\ref{bogo_inv})
this can now be converted to the eigenvectors $\bbox{\Gamma}$.
The dynamical spin correlation function of the spin liquid thus
consists of two components:
First, there is a two-particle continuum,
which dominates for momenta around $(0,0)$. 
Second, there is a single-particle like contribution, which dominates
the cross section
for momentum transfers near $(\pi,\pi)$, and hence should be
identified with the excitations seen in neutron scattering
around this momentum. 
The situation is
the same for ladders, where the single-particle spectrum
is observable in the channel with momentum transfer
perpendicular to the ladder, $k_\perp=\pi$, and the
two-particle continuum for $k_\perp=0$\cite{lad}.\\
Next, we discuss the relationship between condensation
of triplets and antiferromagnetic ordering.
The operator of
staggered magnetization (which is a {\em vector}) can be written as
\begin{eqnarray}
\bbox{M}_s &=& \sqrt{2N}
\frac{\kappa}{4}\sqrt{\frac{n}{n_1}}\sum_{\alpha=x,y}
(\bar{\bbox{\tau}}_{\bbox{Q},\alpha}^\dagger + 
\bar{\bbox{\tau}}_{\bbox{Q},\alpha} )
\nonumber \\
&=& \sqrt{2N} \frac{\kappa}{4}\sqrt{\frac{n}{n_1}}
\sqrt{2+\chi(2)} (\bbox{\tau}_{\bbox{Q},+}^\dagger 
+ \bbox{\tau}_{\bbox{Q},+} ).
\end{eqnarray}
Here we have used the fact that for the high-symmetry
momentum $\bbox{Q}$ the overlap and Hamilton matrix are
trivially diagonalized by the symmetric and antisymmetric
combinations of the bonds in $x$ and $y$-direction ,
$\bbox{\tau}_{\bbox{Q},\pm}^\dagger= \frac{1}{\sqrt{2}}
(\bar{\bbox{\tau}}_{\bbox{Q},x}^\dagger \pm 
\bar{\bbox{\tau}}_{\bbox{Q},y}^\dagger)$.
Then, introducing the
3D unit vector $\bbox{\Omega}$, we can construct the coherent state
\[
|\Psi_\lambda \rangle =
 e^{\lambda \sqrt{N} \bbox{\Omega}\cdot
\bbox{\tau}_{\bbox{Q},+}^\dagger} |RVB \rangle.
\]
This state corresponds to a condensate of the
$\bbox{\tau}_{\bbox{Q},+}^\dagger$ Bosons, whose density
is given by $\delta_{\bbox{Q}}=\lambda^2$.
The staggered magnetization per site is
\[
\bbox{m}_s = \frac{\kappa}{2}\;
\sqrt{ \frac{\delta_{\bbox{Q}} n (1+\chi(2)/2)}{n_1}}\;  \bbox{\Omega}.
\]
Inserting the value for the condensate
density $\delta_{\bbox{Q}}$ obtained from (\ref{conden}) we 
obtain the value $m_s=0.25$. Most current estimates
for the 2D Heisenberg 
antiferromagnet are around $m_s=0.35$\cite{manousakis}.\\
As mentioned above, the identification of
antiferromagnetic ordering and condensation of
some kind of effective Bosons is quite reminiscent
of the treatment of Hirsch and Tang\cite{hirschtang} in the
framework of Schwinger-Boson mean-field theory\cite{arovasauerbach}.
In the present theory the
direction of the staggered magnetization is given by
the unit vector $\bbox{\Omega}$, which can be chosen
arbitrarily. Condensation of the triplet Bosons
determines the total density of triplets, but not their
distribution over the three spin species.
The `ground state' thus is actually an entire manifold of states
which can be transformed into one another via SO(3) rotations of the
vector $\bbox{\Omega}$. One might thus conjecture
the existence of low-energy states where the direction of
$\bbox{\Omega}$ changes slowly in real space. These states,
which corresponds to a slow fluctuation of the antiferromagnetic
order parameter may be describable in terms of a nonlinear $\sigma$-model
- we defer a detailed discussion of this issue to a separate paper.\\
To conclude this section, we discuss the relationship
with Zhang's SO(5) symmetric theory of cuprate 
superconductivity\cite{zhang}.
The preceding discussion has shown that an antiferromagnetic state
can be viewed as a spin liquid where the triplet-like bond Bosons
have condensed into the state which has momentum $(\pi,\pi)$ and $s$-like
symmetry under point group operations. The antiferromagnetic order
parameter, a real 3-vector, then is the vector of
condensation amplitudes of the three bond-Boson species.
This interpretation of the antiferromagnetic state in fact is
a key ingredient for a microscopic interpretation of Zhang's
SO(5) theory of cuprate superconductor. Namely the
$\pi$-operator, which acts as the `ladder operator in charge direction'
in Zhang's representation of the SO(5) angular momentum algebra\cite{zhang},
precisely converts an $s$-like combination
of bond triplets with momentum
$(\pi,\pi)$ into a nearest neighbor $d$-wave hole pair.
This can be seen by writing the $\pi$-operator in real space:
\begin{eqnarray}
\pi_z &=& \sum_i e^{i \vec{Q} \cdot \vec{R}_i}
[\; 
(\;c_{i,\uparrow} c_{i+\hat{x},\downarrow} +
c_{i,\downarrow} c_{i+\hat{x},\uparrow}\;)
\nonumber \\
&\;&\;\;\;\;\;\;\;\;\;\;\;\;\;\;\;\;\;\;
- (\;c_{i,\uparrow} c_{i+\hat{y},\downarrow} +
c_{i,\downarrow} c_{i+\hat{y},\uparrow}\;)\;].
\end{eqnarray}
Introducing a further bond Boson $h^\dagger$, which stands for the
hole pair, this could be written as
\[
\bbox{\pi} = h_{\bbox{K},x}^\dagger\bar{\bbox{\tau}}_{\bbox{Q},x}^{} 
- h_{\bbox{K},y}^\dagger\bar{\bbox{\tau}}_{\bbox{Q},y}^{},
\]
where the momentum $\bbox{K}=(0,0)$.
Acting with 
$(\bbox{\Omega} \cdot \bbox{\pi})^{\rho N}$ onto an undoped state
with  the triplet condensate densities $\rho\bbox{\Omega}$
will therefore
convert this state into a condensate of $d$-like nearest neighbor hole-pairs
with momentum $(0,0)$.
Viewed in this way, the connection between the antiferromagnetic and the
superconducting state appears quite natural. The only drawback
would be that the hole pairs are strictly nearest-neighbor
pairs, i.e. the present version of the $\pi$-operator
neglects charge fluctuations.
\section{Discussion}
In summary, we have discussed the excitation spectrum
of a completely disordered and homogeneous spin system.
Thereby we used an approach which might be viewed as a generalization
of spin wave theory: whereas spin wave theory describes
fluctuations around a N\'eel ordered `vacuum', we have instead
used the nearest neighbor RVB state as basis for self-consistently
constructing the excitation spectrum.
This general idea of treating a completely disordered state
is probably more widely applicable, for example
to treat the `orbital liquids' proposed
recently\cite{feiner} for manganates.\\
As a first key result, we then found that the elementary excitations
of the singlet soup-like vacuum are not Fermionic `spinons',
but rather Bosonic excitations $\bbox{\tau}_i$, which
can be viewed as excited dimers propagating through the system
while constantly resonating between $x$ and $y$ direction of the dimer.
Following a similar procedure as in the Zhang-Rice derivation
of the t-J model, we could write down a Hamiltonian for these
Bosonic excitations. The Hamiltonian contains an energy of formation
for the triplets, a term describing their propagation
and a term describing pair creation and annihilation.
All quantities in the derivation are quite well-defined,
and can in principle be obtained by numerical techniques.
In the present work we have actually attempted an `ab initio
calculation' of the various parameters in the effective
Hamiltonian - a promising alternative to this somewhat
clumsy approach might be a semi-empirical
approach, where the parameters in the Hamiltonian are
adjusted to match e.g. experimental data.\\
The obtained ground state is an exact spin singlet, translationally
invariant and isotropic - in short, it has precisely the symmetry properties 
expected for a spin liquid.
The elementary excitations is a triplet mode, which
reaches its lowest energy at $\bbox{Q}$$=$$(\pi,\pi)$.
Such a triplet mode is the most natural generalization
of an antiferromagnetic spin-wave to the spin-liquid state.
Condensation of these bond Bosons into the state with momentum
$(\pi,\pi)$ then corresponds to antiferromagnetic ordering
of the system.
The breaking of rotational symmetry in spin space is due to
fixing (different) condensation amplitudes
for the three components of the triplet mode. The latter
correspond to the three possible
components of the antiferromagnetic order parameter.\\
In the present work we have chosen the
completely isotropic nearest neighbor RVB-state as the starting point
for constructing the triplet-Hamiltonian.
This is appropriate for an truly isotropic system, 
which probably is realized at finite doping and/or sufficiently
high temperature. Using a translationally invariant state, however,
is not mandatory. For example by redefining the operator $S$ as
\begin{equation}
S = \sum_i ( (1+\epsilon) s_{i,i+\hat{x}}^\dagger +  
(1-\epsilon) s_{i,i+\hat{y}}^\dagger),
\label{sdef1}
\end{equation}
with $\epsilon>0$ we can obviously generate a `singlet soup' with a
preference for singlet orientation in $x$-direction.
It is tempting to speculate that this may be
appropriate for orthorhombic La$_{1-x}$Sr$_x$CuO$_4$,
where at $x=0.125$ columnar order is known to
exist. Analogous calculation as above, but with
a finite $\epsilon$ may thus be appropriate to discuss the
excitation spectrum of this material.\\
Finally we note that the present scenario for the excitation spectrum
of the spin-liquid provides a corner stone for a microscopic
interpretation of Zhang's SO(5) symmetric theory
of cuprate superconductors\cite{zhang}. As discussed above, an
antiferromagnetic state may be thought of as being generated
by condensing triplet-like spin excitations of the RVB
spin-liquid into the state with momentum $(\pi,\pi)$.
The antiferromagnetic order parameter corresponds to the vector of
condensation amplitudes of the three triplet species.
Then, SO(5) symmetry  states that such a
triplet-excitation with momentum $(\pi,\pi)$ is
`dynamically equivalent' to a $d_{x^2-y^2}$ hole pair\cite{so5lad}. In fact the
$\pi$-operator\cite{zhang}, which plays the role of a ladder operator
for `rotations in charge-direction'
in the SO(5) theory\cite{zhang}, precisely
replaces a bond-triplet with momentum $(\pi,\pi)$
by a hole pair with momentum $(0,0)$.
The fact that the $\pi$-operator is an approximate
eigenoperator of the Hamiltonian, $[H,\pi]=\omega_0 \pi$,
then  implies that the triplet and the hole pair
are dynamically indistinguishable and differ only by
their energy of formation.
The $\pi$-operator thus would convert a condensate of
triplets into a condensate of hole pairs, and,
by the dynamical equivalence of these two objects,
thus converts the antiferromagnetic ground state at half-filling into
a superconducting state at finite-doping one.
The somewhat `toy-model-type' discussion of Ref.\cite{so5lad} 
thus may very well be transferable almost
literally to the fully planar t-J model.\\
In the present work we have restricted ourselves to
a pure spin system. One my expect, however, that all considerations
go through also for the doped system, with the sole difference
that we get additional renormalizations of the matrix elements
in the effective Hamiltonian due to the fact that
the doped holes reduce the volume available for pair generation
and propagation of triplets. Moreover, we will need terms which describe
the coupling of the triplet branch to the mobile holes.
In the simpler case of hole motion in a spin ladder this program has in fact
already been carried out\cite{lad,sushkov}, leading to quite
satisfactory agreement with numerical results.
For the planar case we defer this to future work.\\
I would like to thank W. Hanke, O. P. Sushkov and
Shou-Cheng Zhang for instructive discussions and helpful
comments.
\section{Appendix}
In this appendix we discuss the numerical work needed to obtain
the values of $\chi(u)$, $\epsilon_0(u)$ and the $\eta$.
To that end all possible dimer coverings of
an $M\times M$ cluster with periodic boundary conditions
are generated on the computer 
the sums in  (\ref{over-real}) are evaluated
numerically. Restrictions on computer memory and CPU time
do not allow to use $M>6$, so that in practice only
$M=4,6$ are possible. 
Tu study the size dependence, we consider the quantities
$\bar{\chi}(L) = \frac{n_1}{n}\chi(L)$; these give the ratio of the
norm of an RVB state covering the exterior of a loop
of length $L$ to the norm of an RVB state covering the entire cluster.
Similarly, we define $\bar{\epsilon}_0(L) = \frac{n_1}{n}\epsilon_0(L)$,
which gives the gain or loss in energy due to a fixed loop of length
$L$. The values for the different
$\bar{\chi}$ and $\bar{\epsilon}_0$ are given in Table \ref{tab1}
and actually show already a quite remarkable independence of
systems size. 
Moreover, the expectation value of the energy/bond,
$E_0/N$, is $-0.334318$ for $M=4$, $-0.313763$ for $M=6$; the estimate for
$L=\infty$ is $-0.302$\cite{LiangDoucotAnderson}. All in all
the data suggest that already the $4\times 4$ cluster gives reasonably 
accurate estimates for the different parameters.\\
Moreover, the data show that the $\bar{\chi}(L)$ and $\bar{\epsilon}(L)$ 
decrease quite
rapidly with $L$. truncation of the series
after $L=2$ thus seems to be quite a reasonable approximation as well.
\begin{table}
\begin{tabular}{l| l l| l l}
 $\;$ & M=2 & $\;$ & M=4 $\;$ \\
\hline
 $L$ & $\bar{\chi}(L)$ & $\bar{\epsilon}_0(L)$ & $\bar{\chi}(L)$ &  
$\bar{\epsilon}_0(L)$\\
\hline
 1   & 0.10953 &  0.13844 & 0.121483 &  0.15142 \\
 2   & 0.03602 &  0.01675 & 0.043671 &  0.01998 \\
 3   & 0.00780 & -0.00197 & 0.012879 & -0.00001 \\
\end{tabular}
\caption{The values of $\bar{\chi}(L)$ and $\bar{\epsilon}_0(L)$
for the two different L.}
\label{tab1}
\end{table}
 
\end{multicols}
\end{document}